\documentclass[conference]{IEEEtran}
\usepackage{amsmath,amsfonts,amssymb}
\usepackage{algorithmic}
\usepackage{algorithm}
\usepackage{array}
\usepackage{caption}
\usepackage{float}
\usepackage{textcomp}
\usepackage{stfloats}
\usepackage{url}
\usepackage{enumerate}
\usepackage{verbatim}
\usepackage{graphicx}
\usepackage{cite}
\usepackage{graphicx}
\usepackage{textcomp}
\usepackage{xcolor}
\usepackage{bm}
\usepackage{subcaption}

\usepackage{subfloat}
\usepackage{balance}
\usepackage{url}
\usepackage{caption}
\usepackage{cite}
\usepackage {epstopdf}
\usepackage{orcidlink} %
\hypersetup{hidelinks} 

\newcommand{\avector}[2]{\boldsymbol{a}_{#1}^{#2}(\varphi_{#1}^e,\varphi_{#1}^a,\tilde{\boldsymbol{#2}})}

\begin{document}
\title{\LARGE Joint Transceiver Design for RIS Enhanced Dual-Functional Radar-Communication with Movable Antenna\vspace{-3mm}}
\author{Ran Yang$^{\dagger}$, Zheng Dong$^{\star}$, Peng Cheng$^{\ddagger}$, Yue Xiu$^{\dagger}$, Yan Li$^{\dagger}$, and Ning Wei$^{\dagger}$\\
	$^{\dagger}$National Key Laboratory of Wireless Communications, UESTC, Chengdu, China\\
	$^{\star}$School of Information Science and Engineering, Shandong University, Qingdao, China\\
	$^{\ddagger}$Department of Computer Science and Information Technology, La Trobe University, Bundoora, Australia\\
	Emails: zhengdong@sdu.edu.cn, p.cheng@latrobe.edu.au, xiuyue12345678@163.com, \{wn, yan.li\}@uestc.edu.cn\vspace{-2mm}}
	



\maketitle
\thispagestyle{empty}
\vspace{-10mm}
\begin{abstract}
Movable antennas (MAs) have shown significant potential in enhancing the performance of dual-functional radar-communication (DFRC) systems. In this paper, we investigate the MA-based transceiver design for DFRC systems, where a reconfigurable intelligent surface (RIS) is employed to enhance the communication quality in dead zones. To enhance the radar sensing performance, we formulate an optimization problem to maximize the radar signal-to-interference-plus-noise ratio (SINR) by jointly optimizing the  beamforming vectors, receiving filter, antenna positions, and RIS reflecting coefficients. To tackle this challenging problem, we develop  a fractional programming-based optimization framework, incorporating block coordinate descent (BCD), successive convex approximation (SCA), and penalty techniques. Simulation results demonstrate that the proposed method can significantly improve the radar SINR, and achieve satisfactory balance between the radar and communication performance compared with existing benchmark schemes.
\end{abstract}


\vspace{-2mm}
\section{Introdution}
\IEEEPARstart{F}{uture} sixth generation (6G) wireless networks are expected to enable vertical applications such as  auto-driving, industrial automation (IA), and virtual reality (VR), which demand higher sensing precision and more reliable communiaction quality\cite{8999605}. These demands necessitate a paradigm shift in existing wireless networks. In particular, dual-functional radar-communication (DFRC) systems, which integrate radar sensing and communication functions into a single platform, have emerged as a promising  solution for the forthcoming 6G era. 
DFRC systems share spectrum resources, hardware facilities, and signal-processing modules, making them highly efficient. Consequently, the DFRC technology has been deployed in various wireless systems, such as   orthogonal time frequency space modulation (OTFS) \cite{yuan2022orthogonal}, non-orthogonal multiple access (NOMA) \cite{mu2022noma}, and unmanned aerial vehicle (UAV)\cite{9293257}.

Despite extensive research on DFRC\cite{8999605,yuan2022orthogonal,mu2022noma,9293257},  the majority of existing studies concentrate on conventional multiple-input multiple-output (MIMO) systems with fixed-position antennas (FPAs). This limits the full exploration of diversity and spatial multiplexing gains of wireless channels, as the channel variations across the continuous spatial field are not completely utilized. Additionally, the fixed geometric configurations of conventional FPA arrays can result in array gain loss during radar beamforming tasks. 

Recently, movable antennas (MAs), have been proposed to overcome the fundamental limitations of conventional FPA-based systems. These MAs allow antenna elements to be flexibly adjusted within a local moving region\cite{2024arXiv240715448N}. In the MA-assisted systems, each antenna element is connected to a radio frequency (RF) chain via flexible cables to support active movement\cite{10508218}. Recent studies have shown that properly adjusting the positions of MAs within a designated area can significantly boost dual-task performance. For instance, the sum of the communication rate and sensing mutual information (MI) was maximized in \cite{lyu2025movable} by jointly optimizing the antenna coefficients and positions. In \cite{zhou2024fluid}, the communication rate was maximized while meeting sensing beampattern gain requirements.  Although the aforementioned research has demonstrated the superiority of MA over FPA, the susceptibility of radio signals to blockage and attenuation events has not been well-investigated, which inevitably limits the potential of the MA technology.

On the other hand, the reconfigurable intelligent surface (RIS), has drawn wide attention for its capability to provide additional links when the direct link is blocked or in deep fading\cite{9847080}. To deal with the blockage and attenuation issues, several research has explored the integration of RIS into the MA-assisted DFRC systems  \cite{ma2024movable,wu2024movable}. In \cite{ma2024movable}, the authors investigated the maximization of the minimum beampattern gain for an MA aided RIS-DFRC system. Furthermore, the work in \cite{wu2024movable} proposed an MA-based secure transmission scheme against eavesdropping. However, these studies  mainly concentrated on  designing the transmitter \cite{ma2024movable} or considering the clutter-free environment\cite{wu2024movable},  which generally leads to the degraded DRFC performance.  To fully exploit the degrees of freedom (DoFs) in channel reconfiguration enabled  by the MA technology, a general MA-based transceiver design framework for the RIS-enhanced DFRC systems is greatly desired.

Motivated by the above observations, we investigate the joint   transceiver design for RIS-enhanced DFRC systems with movable antenna. More specifically, we aim to maximize the radar signal-to-interference-plus-noise ratio (SINR) subject to the constraints of communication quality of service (QoS). In particular, our contributions can be summarized as follows:
\begin{itemize}
	\item We develop an optimization framework of the MA-based transceiver design for RIS-enhanced DFRC systems. An optimization problem is formulated  to maximize the radar SINR by joint designing the transmitting beamforming, receiving filter, RIS coefficients, as well as positions of the transceiver antennas.
	\item By taking advantage of  the fractional programming (FP) method, we show that the resulting non-convex problem can be successfully solved by jointly using the block coordinate descent (BCD), successive convex approximation (SCA), and the penalty technique.
	\item We show that the proposed optimization algorithm can substantially enhance the radar SINR, and achieve a satisfactory trade-off between the radar performance and communication quality compared to the benchmark schemes.
\end{itemize}

\section{System Model}
\begin{figure}[t]
	\centerline{\includegraphics[width=2.3in]{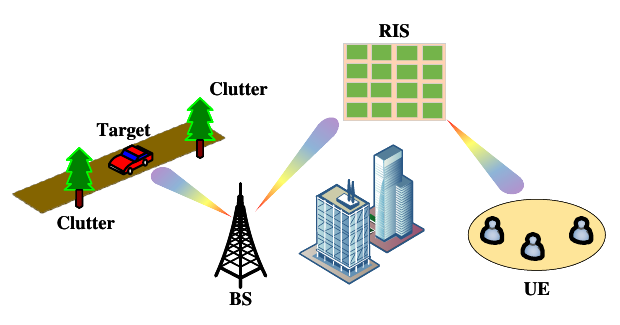}}
	\captionsetup{font={small},labelsep=period,singlelinecheck=off}
	\caption{System model.}
	\label{model}
\end{figure}
 We consider a DFRC system as depicted in Fig. \ref{model}, where the base station (BS) equipped with planar arrays of $N$ transmitting/receiving movable antennas serves $K$ single-antenna users while detects a point-like target in the presence of $Q$ clutters. It is assumed that the direct links from the BS to the users are not available due to blockages, thus an $M$-element RIS is deployed to creat virtual line-of-sight (LoS) links. The 2D moving regions for the transceiver antennas are respectively denoted by $\mathcal{C}_t$ and $\mathcal{C}_r$, both consisting of  a square region of size $A\times A$. Moreover, the positions of the $n$-th transmitting and  receiving MAs are denoted by $\boldsymbol{t}_n = [x_n^t,y_n^t]^T$ and $\boldsymbol{r}_n = [x_n^r,y_n^r]^T$, with the reference points for the regions $\mathcal{C}_t$ and $\mathcal{C}_r$ represented by $\boldsymbol{o}^t = \boldsymbol{o}^r = [0,0]^T$, respectively. 
 \vspace{-2mm}
 \subsection{Communication Model}


Let $\boldsymbol{w}_k$ represent the beamforming vector for the user $k$, the signal $\boldsymbol{x}\in \mathbb{C}^{N\times 1}$ transmitted by the DFRC BS can be written as
\begin{equation}
	\boldsymbol{x} =\boldsymbol{Ws} = \sum_{k=1}^{K}\boldsymbol{w}_ks_k,
\end{equation} 
where $\boldsymbol{s} = [s_1,\dots,s_K]^T\in \mathbb{C}^{K\times1}$ with $\mathbb{E}[\boldsymbol{s}\boldsymbol{s}^H] = \boldsymbol{I}_K$ represents the data symbols for all the users\footnote{We assmue that the BS generates communication signals only to perform radar sensing, which requires least changes in existing wireless networks.}. Given that the signal propagation distance is significantly larger than the size of the moving region, the far-field condition is met. Consequently,  the far-field response model can be applied for channel modeling\cite{10243545}. Specifically, the  angle-of-arrival (AoA), angle-of-departure (AoD), and amplitude of the complex coefficient for each link remain constant despite the movement of the MAs. Note that we adopt the geometric model for the communication channels, thus the number of transmission paths at different nodes are the same, denoted by $L$\cite{10318061}. The elevation and azimuth angles of the $j$-th transmission path  at the BS and RIS are given by  $\psi^e_j \in [0,\pi],\psi^a_j\in [0,\pi]$ and $\phi^e_j \in [0,\pi],\phi^a_j\in [0,\pi]$, respectively. Then, for the $j$-th transmission path, the signal propagation difference between the position of the $n$-th transmitting MA $\boldsymbol{t}_n$ and the reference point $\boldsymbol{o}^t$ is given by
\begin{equation}\label{rho}
\rho(\boldsymbol{t}_n,\psi^e_j,\psi_j^a) = x^t_n\sin\psi^e_j\cos\psi^a_j+y^t_n\cos\psi^e_j.
\end{equation}  
Consequently, the field response vector (FRV) at $\boldsymbol{t}_n$ is given by
\begin{equation}
\boldsymbol{g}(\boldsymbol{t}_n) = \left[e^{\jmath\frac{2\pi}{\lambda}\rho(\boldsymbol{t}_n,\psi^e_1,\psi_1^a)},\dots,e^{\jmath\frac{2\pi}{\lambda}\rho(\boldsymbol{t}_n,\psi^e_{L},\psi_{L}^a)}\right]^T \in \mathbb{C}^{L \times 1},
\end{equation} 
where $\lambda$ is the carrier wavelength. Therefore,  the field response matrix (FRM) of the BS-RIS link for all $N$ transmitting MAs is given by
\begin{equation}
	\boldsymbol{G}(\tilde{\boldsymbol{t}})\triangleq\left[\boldsymbol{g}(\boldsymbol{t}_1),\boldsymbol{g}(\boldsymbol{t}_2),\dots,\boldsymbol{g}(\boldsymbol{t}_N)\right]\in \mathbb{C}^{L\times N},
\end{equation}
where $\tilde{\boldsymbol{t}}\triangleq\left[\boldsymbol{t}_1^T,\boldsymbol{t}_2^T,\dots,\boldsymbol{t}_N^T\right]^T\in \mathbb{R}^{2N\times 1}$. Similarly, the FRV at the $m$-th RIS reflecting element can be derived as 
\begin{equation}
	\boldsymbol{f}(\boldsymbol{s}_m) = \left[e^{\jmath\frac{2\pi}{\lambda}\rho(\boldsymbol{s}_m,\phi_1^e,\phi_1^a)},\dots,e^{\jmath\frac{2\pi}{\lambda}\rho(\boldsymbol{s}_m,\phi_{L}^e,\phi_{L}^a)}\right]^T\in \mathbb{C}^{L\times1},
\end{equation}
where $\boldsymbol{s}_m = [x_m^s,y_m^s]^T$ is the coordinate of the $m$-th element, and $\rho(\boldsymbol{s}_m,\phi_j^e,\phi_j^a)$ denotes the propagation distance difference  for the $j$-th path between the position $\boldsymbol{s}_m$ and origin of the RIS,  $\boldsymbol{o}^s = [0,0]^T$. Then, the FRM at the RIS can be given by
\begin{equation}
	\boldsymbol{F}({\tilde{\boldsymbol{s}}}) \triangleq \left[\boldsymbol{f}(\boldsymbol{s}_1),\boldsymbol{f}(\boldsymbol{s}_2),\dots,\boldsymbol{f}(\boldsymbol{s}_M)\right]^T\in \mathbb{C}^{L\times M}.
\end{equation}
Let $\boldsymbol{\Sigma}=\mathrm{diag}\{\sigma_{1,1},\sigma_{2,2},\dots,\sigma_{L,L}\}\in \mathbb{C}^{L\times L}$ denote the path response matrix (PRM) of the BS-RIS link, the channel matrix can be expessed as 
\begin{equation}\label{channel}
	\boldsymbol{H}(\tilde{\boldsymbol{t}}) = \boldsymbol{F}({\tilde{\boldsymbol{s}}})^H\boldsymbol{\Sigma}\boldsymbol{G}(\tilde{\boldsymbol{t}}).
\end{equation}
Assuming that all users are equipped with a single fixed antenna, the  channel $\boldsymbol{h}_k\in \mathbb{C}^{M\times 1}$ between the RIS and the $k$-th user can be derived in a similar fashion as  (\ref{channel}) and is thus omitted for brevity. The received signal of the $k$-th user is given by
\begin{equation}
	y_k = \underbrace{\boldsymbol{h}_k^H\boldsymbol{V}\boldsymbol{H}(\tilde{\boldsymbol{t}})\boldsymbol{w}_ks_k}_{\text{desired signal}}+\underbrace{\sum_{j=1,j\neq k}^{K}\boldsymbol{h}_k^H\boldsymbol{V}\boldsymbol{H}(\tilde{\boldsymbol{t}})\boldsymbol{w}_js_j}_{\text{inter-user interference}}+n_k,
\end{equation}
where $\boldsymbol{V} = \mathrm{diag}\{[v_1,\dots,v_M]^T\}\in \mathbb{C}^{M\times M}$ with $v_m = e^{\jmath\theta_m}$ is the reflection coefficient matrix, and $n_k \sim \mathcal{CN}(0,\sigma_k^2)$ is the additive white Gaussian noise (AWGN).  
Then, the SINR of the $k$-th user is given by
\begin{equation}
	\Gamma_k(\boldsymbol{W},\boldsymbol{V},\tilde{\boldsymbol{t}}) = \frac{|\boldsymbol{h}_k^H\boldsymbol{V}\boldsymbol{H}(\tilde{\boldsymbol{t}})\boldsymbol{w}_k|^2}{\sum_{j=1,j\neq k}^{K}|\boldsymbol{h}_k^H\boldsymbol{V}\boldsymbol{H}(\tilde{\boldsymbol{t}})\boldsymbol{w}_j|^2+\sigma_k^2},
\end{equation}
\subsection{Radar Model}
We adopt the LoS channel model for the sensing channels between the BS and the target/clutters. Let $\varphi_0^e$ and $\varphi_0^a$ denote the elevation and azimuth angle between the target and the BS, the receiving and transmitting steering vectors can be given by $	\mathbf{a}_0^r(\varphi_0^e,\varphi_0^a,\tilde{\boldsymbol{r}}) = [e^{\jmath\frac{2\pi}{\lambda}\rho(\boldsymbol{r}_1,\varphi_0^e,\varphi_0^a)},\dots,e^{\jmath\frac{2\pi}{\lambda}\rho(\boldsymbol{r}_N,\varphi_0^e,\varphi_0^a)}]^T$  and $\mathbf{a}_0^t(\varphi_0^e,\varphi_0^a,\tilde{\boldsymbol{t}}) = [e^{\jmath\frac{2\pi}{\lambda}\rho(\boldsymbol{t}_1,\varphi_0^e,\varphi_0^a)},\dots,e^{\jmath\frac{2\pi}{\lambda}\rho(\boldsymbol{t}_N,\varphi_0^e,\varphi_0^a)}]^T$, respectively. Let  $\boldsymbol{A}_0(\tilde{\boldsymbol{r}},\tilde{\boldsymbol{t}}) = 	\mathbf{a}_0^r(\varphi_0^e,\varphi_0^a,\tilde{\boldsymbol{r}})\mathbf{a}_0^t(\varphi_0^e,\varphi_0^a,\tilde{\boldsymbol{t}})^H$ denote the response matrix for the target, the received echo signal can be given by
\begin{equation}
	y_r = \alpha_0 \boldsymbol{u}^H\boldsymbol{A}_0(\tilde{\boldsymbol{r}},\tilde{\boldsymbol{t}})\boldsymbol{x} + \boldsymbol{u}^H\sum_{q=1}^{Q}\alpha_q\boldsymbol{A}_q(\tilde{\boldsymbol{r}},\tilde{\boldsymbol{t}})\boldsymbol{x} + \boldsymbol{u}^H\boldsymbol{n}_r,
\end{equation}
where  $\boldsymbol{A}_q(\tilde{\boldsymbol{r}},\tilde{\boldsymbol{t}}) =  \mathbf{a}_q^r(\varphi_q^e,\varphi_q^a,\tilde{\boldsymbol{r}})\mathbf{a}_q^t(\varphi_q^e,\varphi_q^a,\tilde{\boldsymbol{t}})^H$ represents the response matrix for the $q$-th clutter, and $\boldsymbol{u}\in \mathbb{C}^{N\times1}$  is the receiving filter.  Here, $\alpha_0$ and $\alpha_q$ are the complex coefficients including the  radar cross section (RCS) and the cascaded complex gain of the target and $q$-th clutter,  with $\mathbb{E}\{|\alpha_0|^2\} = \zeta_0^2 ~\text{and}~ \mathbb{E}\{|\alpha_q|^2\} = \zeta_q^2$, respectively. Besides, $\boldsymbol{n}_r \sim\mathcal{CN}(0,\sigma_r^2\boldsymbol{I}_N)$ is the AWGN at the BS. Define that $\boldsymbol{\Xi} \triangleq \sum_{q=1}^{Q}\zeta_q^2\boldsymbol{A}_q(\tilde{\boldsymbol{r}},\tilde{\boldsymbol{t}})\boldsymbol{W}\boldsymbol{W}^H\boldsymbol{A}_q(\tilde{\boldsymbol{r}},\tilde{\boldsymbol{t}})^H$,  likewise in \cite{10132501}, the radar SINR is calculated by 
\begin{equation}
	\Gamma_r(\boldsymbol{W},\tilde{\boldsymbol{r}},\tilde{\boldsymbol{t}},\boldsymbol{u})  =  \frac{\zeta_0^2|\boldsymbol{u}^H\boldsymbol{A}_0(\tilde{\boldsymbol{r}},\tilde{\boldsymbol{t}})\boldsymbol{x}|^2}{\boldsymbol{u}^H\left(\boldsymbol{\Xi}+\sigma_r^2\boldsymbol{I}_N\right)\boldsymbol{u}}. 
\end{equation}

To maximize the radar SINR $\Gamma_r(\boldsymbol{W},\tilde{\boldsymbol{r}},\tilde{\boldsymbol{t}},\boldsymbol{u})$, the solution for the filter $\boldsymbol{u}$ can be obtained by solving the minimum variance distortionless response (MVDR) problem\cite{1449208}, i.e., 
\begin{equation}\label{filter}
	\boldsymbol{u}^\star = \beta\left(\boldsymbol{\Xi}+\sigma_r^2\boldsymbol{I}_N\right)^{-1}\boldsymbol{A}_0(\tilde{\boldsymbol{r}},\tilde{\boldsymbol{t}})\boldsymbol{x},
\end{equation}
where $\beta$ is an auxiliary constant. According to \cite{9724174}, the corresponding radar SINR can be further given by
\begin{align} 
	\Gamma_r(\boldsymbol{W},\tilde{\boldsymbol{r}},\tilde{\boldsymbol{t}})=~&{\mathbb{E}}\left [{ {\frac {{{{\zeta_0}^{2}}{{\left |{ {{{\boldsymbol{u}}^{H}}{\boldsymbol{A}_0(\tilde{\boldsymbol{r}},\tilde{\boldsymbol{t}})}{\mathbf{x}}} }\right |}^{2}}}}{{{{\boldsymbol{u}}^{H}}\left ({{\boldsymbol{\Xi} + {\sigma_r^2\boldsymbol{I}_N}} }\right){\boldsymbol{u}}}}} }\right] \notag\\
	\mathop = \limits ^{\left ({a }\right)}~&{\mathbb{E}}\left[ \zeta_0^2\boldsymbol{x}^H\boldsymbol{A}_0(\tilde{\boldsymbol{r}},\tilde{\boldsymbol{t}})^H(\boldsymbol{\Xi}+\sigma_r^2\boldsymbol{I}_N)^{-1}\boldsymbol{A}_0(\tilde{\boldsymbol{r}},\tilde{\boldsymbol{t}})\boldsymbol{x}\right] \notag \\
	\mathop = \limits ^{\left ({b }\right)}~& {\mathrm{tr}}\left({\boldsymbol{\Phi}}\boldsymbol{W}\boldsymbol{W}^H\right),
\end{align} 
where ${\boldsymbol{\Phi}} = \zeta_0^2\boldsymbol{A}_0(\tilde{\boldsymbol{r}},\tilde{\boldsymbol{t}})^H(\boldsymbol{\Xi}+\sigma_r^2\boldsymbol{I}_N)^{-1}\boldsymbol{A}_0(\tilde{\boldsymbol{r}},\tilde{\boldsymbol{t}}).$ Note that the procedure (a) is due to the optimal receiving filter $\boldsymbol{u}^\star$ in (\ref{filter}), and the procedure (b) holds due to $\mathbb{E}[ \boldsymbol{xx}^H] = \boldsymbol{W}\boldsymbol{W}^H$.
\subsection{Problem Formulation}
 Based on the above performance metrics, we focus on the maximization of the radar SINR by jointly designing the beamforming vectors, the positions of the transceiver antennas, the receiving filter, and the RIS reflecting coefficients. Specifically, the optimization problem can be formulated as follows 
\begin{subequations}\label{Problem1}
	\begin{alignat}{2}
		&\underset{ \boldsymbol{W},\boldsymbol{V}, \tilde{\boldsymbol{r}},\tilde{\boldsymbol{t}}}{\max}~ \Gamma_r(\boldsymbol{W},\tilde{\boldsymbol{r}},\tilde{\boldsymbol{t}})\label{P1_objective}\\
		&\,\text {s.t.}~ \Gamma_k(\boldsymbol{W},\boldsymbol{V},\tilde{\boldsymbol{t}}) \ge \gamma_k,\forall k,\label{P1_Com}\\
		&\hphantom {s.t.~}||\boldsymbol{t}_n-\boldsymbol{t}_{n'}||_2^2 \ge D^2,|| \boldsymbol{r}_n-\boldsymbol{r}_{n'}||_2^2 \ge D^2,\forall n\neq n',\label{P1_MA1}\\
		&\hphantom {s.t.~}  \boldsymbol{t}_n \in \mathcal{C}_t,\boldsymbol{r}_n \in \mathcal{C}_r,\forall n,\label{P1_MA2}\\
		&\hphantom {s.t.~}
		\sum_{k=1}^{K}\boldsymbol{w}_k^H\boldsymbol{w}_k \le P_t,\label{P1_Pt}\\
		&\hphantom {s.t.~}
		|{v}_m|^2 = 1, \forall m,\label{P1_RIS}	
	\end{alignat} 
\end{subequations} 
where the constraints in (\ref{P1_Com}) ensure that the SINR at the $k$-th user is no less than the predefined threshold $\gamma_k$, $D$ represents the minimum distance between the MAs to prevent  coupling effects, $P_t$ is the maximum transmission power, and  (\ref{P1_RIS}) is the unit-modulus constraints on RIS. It is challenging to solve (\ref{Problem1}) due to the non-convexity of (\ref{P1_objective}), (\ref{P1_Com}), (\ref{P1_MA1}), (\ref{P1_RIS}), as well as the coupling of the variables. 
\section{Proposed Solution}
In this section, we first reformulate the objective function (\ref{P1_objective}) into a more tractable form by using the fractional programming (FP) technique\cite{8314727}. Then, the block coordinate descent algorithm, incorporating the successive convex approximation and penalty technique, is  proposed to solve the problem (\ref*{Problem1}), the details of which are elaborated as follows.

Specifically, according to the FP technique,  $\Gamma_r(\boldsymbol{W},\tilde{\boldsymbol{r}},\tilde{\boldsymbol{t}})$ can be equivalently rewritten as
\begin{align}\label{objective1}
	\hat{\Gamma}_r(&\boldsymbol{W},\tilde{\boldsymbol{r}},\tilde{\boldsymbol{t}},\boldsymbol{\Lambda})  \notag\\ 
	&\triangleq \zeta_0^2 \mathrm{tr}\left(2\Re\{\boldsymbol{W}^H\boldsymbol{A}_0(\tilde{\boldsymbol{r}},\tilde{\boldsymbol{t}})^H\boldsymbol{\Lambda}\}-\boldsymbol{\Lambda}^H(\boldsymbol{\Xi}+\sigma_r^2\boldsymbol{I}_N)\boldsymbol{\Lambda}\right),
\end{align} 
where $\boldsymbol{\Lambda}\in \mathbb{C}^{N\times K}$ is the auxiliary variable. Based on the reformulated objective function (\ref{objective1}), we now propose a BCD algorithm to obtain an efficient solution for the problem (\ref*{Problem1}).
\subsection{Updating Auxiliary Variable}
We note that the optimization on $\boldsymbol{\Lambda}$ is a unconstrained problem, and the optimal closed-form solution can be directly given by \cite{8314727}
\begin{equation}\label{Auxiliary}
	\boldsymbol{\Lambda}^\star =\left (\boldsymbol{\Xi}+\sigma_r^2\boldsymbol{I}_N\right)^{-1}\boldsymbol{A}_0(\tilde{\boldsymbol{r}},\tilde{\boldsymbol{t}})\boldsymbol{W}.
\end{equation}
\subsection{Updating Transmitting Beamforming}
With $\tilde{\boldsymbol{r}},\tilde{\boldsymbol{t}},\boldsymbol{\Lambda}$ and $\boldsymbol{V}$ fixed, the problem (\ref{Problem1}) is simplified as follows: 
\begin{subequations}\label{beamforming}
	\begin{alignat}{2}
		&\underset{ \boldsymbol{W}}{\max} ~~ \hat{\Gamma}_r(\boldsymbol{W})\\
		&\,\text {s.t.}~ \Gamma_k(\boldsymbol{W}) \ge \gamma_k,\forall k,\label{beamforming_Com}\\
		&\hphantom {s.t.~}
		\sum_{k=1}^{K}\boldsymbol{w}_k^H\boldsymbol{w}_k \le P_t.\label{beamforming_Pt}
	\end{alignat} 
\end{subequations}
The non-convexity of (\ref{beamforming}) lies in $|\boldsymbol{h}_k^H\boldsymbol{V}\boldsymbol{H}(\tilde{\boldsymbol{t}})\boldsymbol{w}_k|^2$ of (\ref*{beamforming_Com}). Here, we employ the SCA method to deal with this issue\cite{razaviyayn2014successive}. To elaborate, we approximate $|\boldsymbol{h}_k^H\boldsymbol{V}\boldsymbol{H}(\tilde{\boldsymbol{t}})\boldsymbol{w}_k|^2$   with its global under-estimator by using the first-order Taylor expansion, i.e., 
\begin{align}
	|\boldsymbol{h}_k^H\boldsymbol{V}\boldsymbol{H}(\tilde{\boldsymbol{t}})\boldsymbol{w}_k|^2 \ge  2\Re\{(\boldsymbol{w}_k&^{(l)})^H\tilde{\boldsymbol{H}}_k(\tilde{\boldsymbol{t}})\boldsymbol{w}_k\}\notag\\
	&-(\boldsymbol{w}_k^{(l)})^H\tilde{\boldsymbol{H}}_k(\tilde{\boldsymbol{t}})\boldsymbol{w}_k^{(l)},
\end{align}
where $\tilde{\boldsymbol{H}}_k(\tilde{\boldsymbol{t}}) =\boldsymbol{H}(\tilde{\boldsymbol{t}})^H\boldsymbol{V}^H\boldsymbol{h}_k \boldsymbol{h}_k^H\boldsymbol{V}\boldsymbol{H}(\tilde{\boldsymbol{t}})$, and $\boldsymbol{w}_k^{(l)}$ is the obtained solution in the $l$-th iteration. Thus, the constraints (\ref{beamforming_Com}) can be approximated by 
\begin{align}
	2\Re\{(\boldsymbol{w}_k^{(l)})^H\tilde{\boldsymbol{H}}_k&(\tilde{\boldsymbol{t}})\boldsymbol{w}_k\}-(\boldsymbol{w}_k^{(l)})^H\tilde{\boldsymbol{H}}_k(\tilde{\boldsymbol{t}})\boldsymbol{w}_k^{(l)}\notag\\ &\ge \gamma_k(\sum_{j=1,j\neq k}^{K}\boldsymbol{w}_j^H\tilde{\boldsymbol{H}}_k(\tilde{\boldsymbol{t}})\boldsymbol{w}_j+\sigma_k^2),\forall k.
\end{align}
Thus, the problem (\ref{beamforming}) is convex and can be solved by off-the-shelf solvers, e.g., the CVX tool\cite{boyd2004convex}.
\subsection{Updating Reflecting Coefficients}
Now we carry out optimization on RIS coefficients with $\boldsymbol{W},\boldsymbol{r},\boldsymbol{t},\boldsymbol{\Lambda}$ fixed.  Note that the objective function (\ref{objective1}) is independent of $\boldsymbol{V}$, which indicates that the design of $\boldsymbol{V}$ is a feasibility-check problem and the solution will not directly affect  (\ref*{objective1}). Therefore, in order to provide additional DoFs for optimization on other variables, we propose to maximize the lower bound of the communication SINR by updating $\boldsymbol{V}$\cite{9847080}. Specifically, we introduce a slack variable $\eta$ to refomulate the problem (\ref{Problem1}) as
\addtolength{\topmargin}{0.042in}
	\begin{subequations}\label{RIS1}
	\begin{alignat}{2}
		&\underset{ \boldsymbol{V},\eta}{\max} \quad \eta\label{RIS1_objective}\\
		&\,\text {s.t.}~ \Gamma_k(\boldsymbol{V})\ge \eta,\forall k,\label{RIS1_Com}\\
		&\hphantom {s.t.~}
	|{v}_m|^2 = 1,\forall m.\label{RIS1_RIS}	
	\end{alignat} 
\end{subequations} 
The problem (\ref{RIS1}) is intractable due to the non-convexity of (\ref{RIS1_Com}) and (\ref*{RIS1_RIS}). We first deal with the constraints (\ref{RIS1_Com}). By introducing a slack variable $\boldsymbol{z} = [z_1,\dots,z_k]^T\in \mathbb{C}^{K \times 1}$, the  constraints (\ref{RIS1_Com}) can be converted to 
	\begin{subequations}\label{RIS1_Com_FP}
	\begin{alignat}{2}
		&|\boldsymbol{h}_k^H\boldsymbol{V}\boldsymbol{H}(\tilde{\boldsymbol{t}})\boldsymbol{w}_k|^2 \ge  \eta z_k,\forall k,\label{RIS1_Com_FP1}\\
		&\sum_{j=1,j\neq k}^{K}|\boldsymbol{h}_k^H\boldsymbol{V}\boldsymbol{H}(\tilde{\boldsymbol{t}})\boldsymbol{w}_j|^2+\sigma_k^2\le z_k, \forall k\label{RIS1_Com_FP2}.
	\end{alignat} 
\end{subequations} 
Note that the equivalence between (\ref{RIS1_Com}) and (\ref{RIS1_Com_FP}) can be verified by contradiction. However, the constraints (\ref{RIS1_Com_FP1}) are still non-convex. To handle this issue, we first define that $\boldsymbol{v} \triangleq [v_1,v_2,\dots,v_m]^H,\tilde{\boldsymbol{h}}_{k,j} \triangleq \mathrm{diag}(\boldsymbol{h}_k^H)\boldsymbol{H}(\tilde{\boldsymbol{t}})\boldsymbol{w}_j$ and $\tilde{\boldsymbol{H}}_{k,j} \triangleq \tilde{\boldsymbol{h}}_{k,j}\tilde{\boldsymbol{h}}_{k,j}^H$, then we rewrite (\ref{RIS1_Com_FP1}) based on the transformation $ \boldsymbol{h}_k^H\boldsymbol{V}=\boldsymbol{v}^H\mathrm{diag}(\boldsymbol{h}_k^H)$ as 
\begin{equation}
	\boldsymbol{v}^H\tilde{\boldsymbol{H}}_{k,k}\boldsymbol{v}\ge \eta z_k,\forall k.\label{RIS1_Com_FP1_reformulated}
\end{equation}
Next, we replace the left-hand-side (LHS) of (\ref{RIS1_Com_FP1_reformulated}) with its lower bound based on the first-order Taylor expasnsion as 
\begin{equation}
	\boldsymbol{v}^H\tilde{\boldsymbol{H}}_{k,k}\boldsymbol{v}\ge 2\Re\{(\boldsymbol{v}^{(l)})^H\tilde{\boldsymbol{H}}_{k,k}\boldsymbol{v}\} - (\boldsymbol{v}^{(l)})^H\tilde{\boldsymbol{H}}_{k,k}\boldsymbol{v}^{(l)},\forall k,
\end{equation}
where $\boldsymbol{v}^{(l)}$ is the solution obtained in the last iteration. For the right-hand-side (RHS) of (\ref{RIS1_Com_FP1_reformulated}), it is not jointly convex with respect to (w.r.t) $\eta$ and $z_k$ but satisfies
\begin{equation}
	\eta z_k\le \frac{1}{2}\left(\frac{z_k^{(l)}}{\eta^{(l)}}\eta^2+\frac{\eta^{(l)}}{z_k^{(l)}}z_k^2\right), \forall k.
\end{equation}
Then  the constraint (\ref{RIS1_Com_FP1}) can be approximated by
\begin{align}
	2\Re\{(\boldsymbol{v}^{(l)})^H\tilde{\boldsymbol{H}}_{k,k}\boldsymbol{v}\} &- (\boldsymbol{v}^{(l)})^H\tilde{\boldsymbol{H}}_{k,k}\boldsymbol{v}^{(l)}  \notag\\
	&\ge \frac{1}{2}\left(\frac{z_k^{(l)}}{\eta^{(l)}}\eta^2+\frac{\eta^{(l)}}{z_k^{(l)}}z_k^2\right),\forall k.\label{RIS1_Com_FP1_approximated}
\end{align}
Now, we move on to tackle the unit-modulus constraints (\ref{RIS1_RIS}) by using the penalty-based technique. Specifically, we reformulate the problem (\ref{RIS1}) to be a penalized version as
\begin{subequations}\label{RIS2}
	\begin{alignat}{2}
		&\underset{ \boldsymbol{v},\eta}{\max} \quad \eta + \rho_1 \left(||\boldsymbol{v}||^2 - M\right)\label{RIS2_objective}\\
		&\,\text {s.t.}~~\text{(\ref{RIS1_Com_FP2})},~ \text{(\ref{RIS1_Com_FP1_approximated})},\\
		&\hphantom {s.t.~}
		|{v}_m|^2 \le 1,\forall m,\label{RIS2_C3}	
	\end{alignat} 
\end{subequations} 
 where $\rho_1$ is a large positive constant which encourages the solution of (\ref{RIS2}) to satisfy (\ref{RIS1_RIS}). Note that the penalty term $\rho_1 \left(||\boldsymbol{v}||^2 - M\right)$ in (\ref{RIS2_objective}) leads to a non-concave objective function, therefore we follow the principle of the SCA method to iteratively approximate (\ref*{RIS2_objective}) by its first-order Taylor expansion. Consequently, the problem (\ref{RIS2}) can be rewritten as
\begin{subequations}\label{RIS3}
	\begin{alignat}{2}
		&\underset{ \boldsymbol{v},\eta}{\max} \quad \eta + \rho_1 \Re\{2(\boldsymbol{v}^{(l)})^H\boldsymbol{v} -(\boldsymbol{v}^{(l)})^H\boldsymbol{v}^{(l)} \} - \rho_1M\\
		&\,\text {s.t.}~~\text{(\ref{RIS1_Com_FP2})},~\text{(\ref{RIS1_Com_FP1_approximated})},~ \text{(\ref{RIS2_C3})}.
	\end{alignat}
\end{subequations} 
The problem is convex and can be solved by using the CVX tool\cite{boyd2004convex}.  A two-layer loop is  adopted for (\ref{RIS1}). In the outer loop, the penalty factor $\rho_1$ is initially set to a small value, e.g., $10^{-2}\sim10^{-1}$ times the RIS size $M$, to find a proper starting point, then updated until sufficiently large. In the inner loop, the problem (\ref{RIS3}) is solved iteratively to update $\boldsymbol{v}$ and $\eta$ with $\rho_1$ fixed. The steps  are summarized in \textit{Algorithm 1}. 

\begin{algorithm}[t]
	\caption{Penalty Optimization on Reflecting Coefficients}
	\begin{algorithmic}[1] \STATE \textbf{Initialize:} set $l=0,\rho_1>0,\tau>1$, and initialize 
		$\boldsymbol{v}^{(l)}$.		
		\REPEAT	
			
		\STATE $\rho_1$ = $\tau \rho_1$.
		
		\REPEAT	
		
		\STATE Update $\boldsymbol{v}^{(l+1)}$ from the problem (\ref{RIS3}).	
		\STATE $l=l+1$.	
		\UNTIL $||\boldsymbol{v}^{(l)} - \boldsymbol{v}^{(l-1)}||^2 \le \xi_1$.
		
		\UNTIL $(||\boldsymbol{v}^{(l)}||^2 - M) \le \xi_2$.
		\RETURN  $\boldsymbol{v}^\star = \boldsymbol{v}^{(l)}$.
	\end{algorithmic} 
\end{algorithm}
\vspace{-2mm}
\subsection{Updating Antenna Positions}
In this subsection, we focus on the optimization on the MAs. We only need to show the design for the transmitting MAs, as the design for the receiver is similar and thus omitted for brevity.

 In order to expose $\tilde{\boldsymbol{t}}$ in (\ref{objective1}), the objective function $\hat{\Gamma}_r(\tilde{\boldsymbol{t}})$ is first transformed into a more tractable form as (\ref*{objective1_rewritten}), shown at the top of next page.
\begin{figure*}[tp]
	\begin{align}
			\hat{\Gamma}_r(\tilde{\boldsymbol{t}}) = ~ 2\zeta_0^2\Re\{\avector{0}{r}^H\boldsymbol{\Lambda}&\boldsymbol{W}^H\avector{0}{t}\} - \notag \\ &\sum_{q=1}^{Q}\zeta_0^2\zeta_q^2\avector{q}{r}^H\boldsymbol{\Lambda}\boldsymbol{\Lambda}^H\avector{q}{r}\avector{q}{t}^H\boldsymbol{W}\boldsymbol{W}^H\avector{q}{t}.
		\label{objective1_rewritten}
	\end{align}
	\vspace{-0.1cm}
	\hrule
	\vspace{-0.5cm}
\end{figure*}
Note that the terms independent of $\tilde{\boldsymbol{t}}$ in (\ref{objective1_rewritten}) are omitted. Let $\boldsymbol{b} = \zeta_0^2 \boldsymbol{W}\boldsymbol{\Lambda}^H\avector{0}{r},c_q = \zeta_0^2 \zeta_q^2\avector{q}{r}^H\boldsymbol{\Lambda}\boldsymbol{\Lambda}^H\avector{q}{r}$, and $\boldsymbol{D} = \boldsymbol{W}\boldsymbol{W}^H$, $\hat{\Gamma}_r(\tilde{\boldsymbol{t}})$ can be further given by
\begin{align}
	\hat{\Gamma}_r(\tilde{\boldsymbol{t}}) = 2\Re\{\boldsymbol{b}^H&\avector{0}{t}\} - \notag\\ &\sum_{q=1}^{Q}c_q\avector{q}{t}^H\boldsymbol{D}\avector{q}{t}.
\end{align}
Thus the problem (\ref*{Problem1}) can be reformulated as 
\begin{subequations}\label{MA1}
	\begin{alignat}{2}
		&\underset{ \tilde{\boldsymbol{t}}}{\max} \quad \hat{\Gamma}_r(\tilde{\boldsymbol{t}})\label{MA1_objective}\\
		&\,\text {s.t.}~ \Gamma_k(\tilde{\boldsymbol{t}}) \ge \gamma_k,\forall k,\label{MA1_Com}\\
		&\hphantom {s.t.~}||\boldsymbol{t}_n-\boldsymbol{t}_{n'}||_2^2 \ge D^2,\forall n\neq n',\label{MA1_MA1}\\
		&\hphantom {s.t.~}  \boldsymbol{t}_n \in \mathcal{C}_t,\forall n,\label{MA1_MA2}	
	\end{alignat} 
\end{subequations} 

The problem (\ref{MA1}) is intractable  due to the objective function (\ref{MA1_objective}), the constraints (\ref{MA1_Com}) as well as (\ref*{MA1_MA1}). To deal with these issues, the SCA method can be applied. 
\vspace{2mm}
\begin{algorithm}[t] 
	\caption{BCD Optimization Algorithm for (\ref{Problem1})}
	\begin{algorithmic}[1]
		\STATE \textbf{Initialize:} set $\psi = 0$, and initialize $\boldsymbol{\Lambda}^{[\psi]},\boldsymbol{W}^{[\psi]},\boldsymbol{V}^{[\psi]},\tilde{\boldsymbol{t}}^{[\psi]}$ and $\tilde{\boldsymbol{r}}^{[\psi]}$.
		\REPEAT
		\STATE Update $\boldsymbol{\Lambda}^{[\psi+1]}$ via  (\ref{Auxiliary});
		\STATE Update $\boldsymbol{W}^{[\psi+1]}$ by solving (\ref{beamforming});
		\STATE Update $\boldsymbol{V}^{[\psi+1]}$ by solving (\ref{RIS3}) via Algorithm 1;
		\STATE  Update $\tilde{\boldsymbol{t}}^{[\psi+1]}$ by solving (\ref{MA2});
		\STATE  Update $\tilde{\boldsymbol{r}}^{[\psi+1]}$ in a similar fashion as $\tilde{\boldsymbol{t}}^{[\psi+1]}$;
		\STATE Let $\psi = \psi+1$;
		\UNTIL The objective value (\ref{objective1}) converges.
		\STATE Return  $\boldsymbol{\Lambda}^\star,\boldsymbol{V}^\star,\boldsymbol{W}^\star,\tilde{\boldsymbol{t}}^\star,\tilde{\boldsymbol{r}}^\star$.
	\end{algorithmic}
\end{algorithm}
\vspace{-2mm}
\subsubsection{SCA for (\ref{MA1_objective})}
According to the second-order Taylor expansion theorem, the objective function (\ref{MA1_objective}) can be lower bounded by a quadratic surrogate  concave function\cite{razaviyayn2014successive}, i.e.,
\begin{equation}\label{MA1_objective_approximation}
	\hat{\Gamma}_r(\tilde{\boldsymbol{t}})\geq \hat{\Gamma}_r(\tilde{\boldsymbol{t}}^{(l)})+\nabla \hat{\Gamma}_r(\tilde{\boldsymbol{t}}^{(l)})^T(\tilde{\boldsymbol{t}}-\tilde{\boldsymbol{t}}^{\left(l\right)})-\frac{\delta_0}{2}(\tilde{\boldsymbol{t}}-\tilde{\boldsymbol{t}}^{\left(l\right)})^T(\tilde{\boldsymbol{t}}-\tilde{\boldsymbol{t}}^{\left(l\right)}),
\end{equation}
where $\tilde{\boldsymbol{t}}^{(l)}$ denotes the positions of MAs obtained in the $l$-th iteration, $\nabla \hat{\Gamma}_r(\tilde{\boldsymbol{t}}^{(l)}) \in \mathbb{C}^{2N \times 1}$   is the gradient vector at $\tilde{\boldsymbol{t}}^{(l)}$, and  $\delta_0$ is a positive real number satisfying $\delta_0 \boldsymbol{I}_{2N}\succeq \nabla^2\hat{\Gamma} _r(\tilde{\boldsymbol{t}}^{(l)})$ with the Hessian matrix $\nabla^2\hat{\Gamma} _r(\tilde{\boldsymbol{t}}^{(l)}) \in \mathbb{C}^{2N \times 2N}$. Note that the calculation of $\nabla \hat{\Gamma}_r(\tilde{\boldsymbol{t}}^{(l)})$ and $\nabla^2\hat{\Gamma}_r(\tilde{\boldsymbol{t}}^{(l)})$ can be performed in a similar fashion as \cite{10243545}, thus omitted for brevity. 
\subsubsection{SCA for Constraints (\ref*{MA1_Com})}Recalling that $	\boldsymbol{H}(\tilde{\boldsymbol{t}}) = \boldsymbol{F}({\tilde{\boldsymbol{s}}})^H\boldsymbol{\Sigma}\boldsymbol{G}(\tilde{\boldsymbol{t}})$,  the constraints (\ref{MA1_Com}) can be rewritten as
\begin{equation}
	\underbrace{\boldsymbol{a}_k^H \boldsymbol{G}(\tilde{\boldsymbol{t}})\boldsymbol{R}_k\boldsymbol{G}(\tilde{\boldsymbol{t}})^H \boldsymbol{a}_k}_{\triangleq f_k(\tilde{\mathbf{t}})} + \gamma_k \sigma_k^2 \leq 0, \forall k, \label{f_k}
\end{equation}
where $\boldsymbol{a}_k = \boldsymbol{\Sigma}^H\boldsymbol{F}(\boldsymbol{s})\boldsymbol{V}^H\boldsymbol{h}_k = [a_{k,1},a_{k,2},\dots,a_{k,L}]^T\in \mathbb{C}^{L\times 1}$ and $\boldsymbol{R}_k = \sum_{j\neq k}^{K}\gamma_k\boldsymbol{w}_j\boldsymbol{w}_j^H - \boldsymbol{w}_k\boldsymbol{w}_k^H\in \mathbb{C}^{N\times N}$. As $f_k(\tilde{\boldsymbol{t}})$ is neither convex nor concave w.r.t. $\tilde{\mathbf{t}}$, we construct a surrogate function that serves as an upper bound of $f_k(\tilde{\boldsymbol{t}})$ based on the second-order Taylor expansion as follows
\begin{equation}\label{MA1_Com_approxiation}
	f_k(\tilde{\boldsymbol{t}}) \leq f_k(\tilde{\boldsymbol{t}}^{\left(l\right)}) + \nabla f_k(\tilde{\boldsymbol{t}}^{\left(l\right)})^T (\tilde{\boldsymbol{t}} - \tilde{\boldsymbol{t}}^{\left(l\right)}) +\frac{\delta_k}{2}(\tilde{\boldsymbol{t}} - \tilde{\boldsymbol{t}}^{\left(l\right)})^T(\tilde{\boldsymbol{t}} - \tilde{\boldsymbol{t}}^{\left(l\right)}), 
\end{equation} 
 where $\nabla f_k(\tilde{\boldsymbol{t}}) $ and $\nabla^2 f_k(\tilde{\boldsymbol{t}})$ denote the gradient vector and the Hessian matrix of $f_k(\tilde{\boldsymbol{t}})$ over $\tilde{\boldsymbol{t}}$, respectively. A positive real number $\delta_k$ is selected to satisfy $\delta_k \boldsymbol{I}_{2N}\succeq \nabla^2 f_k(\tilde{\boldsymbol{t}})$\cite{10243545}. 
\subsubsection{SCA for Constraints (\ref{MA1_MA1})}
For constraints (\ref{MA1_MA1}), since the term $\left\| \boldsymbol{t}_n - \boldsymbol{t}_{n'} \right\|_2^2$ is a convex function w.r.t. $\boldsymbol{t}_n - \boldsymbol{t}_{n'}$, it can lower bounded by its first-order Taylor expansion at the given points $\boldsymbol{t}_n^{\left(l\right)}$ and $\boldsymbol{t}_{n'}^{\left(l\right)}$, i.e, 
\begin{align}
	\left\| \mathbf{t}_n - \mathbf{t}_{n'} \right\|_2^2 \geq 
	-&\left\| \mathbf{t}_n^{(l)} - \mathbf{t}_{n'}^{(l)} \right\|_2^2 + 
	 2( \mathbf{t}_n^{(l)} - \mathbf{t}_{n'}^{(l)})^T \notag\\ 
	&\times ( \mathbf{t}_n - \mathbf{t}_{n'}),~1\leq n \neq n' \leq N. \label{MA1_MA1_approximation}
\end{align}
Thus, the problem (\ref{MA1}) can be approximated as
\begin{subequations}\label{MA2}
\begin{align}
	\underset{\tilde{\boldsymbol{t}}}{\text{max}}~&\hat{\Gamma}_r(\tilde{\boldsymbol{t}}^{\left(l\right)})+\nabla \hat{\Gamma}_r(\tilde{\boldsymbol{t}}^{(l)})^T(\tilde{\boldsymbol{t}}-\tilde{\boldsymbol{t}}^{\left(l\right)})-\frac{\delta_0}{2}(\tilde{\boldsymbol{t}}-\tilde{\boldsymbol{t}}^{\left(l\right)})^T(\tilde{\boldsymbol{t}}-\tilde{\boldsymbol{t}}^{\left(l\right)}) \\
	\text{s.t.}
	&~\boldsymbol{t}_n\in \mathcal{C}_t,~1\leq n \leq N,\\		
	&~-\left\| \boldsymbol{t}_n^{(l)} - \boldsymbol{t}_{n'}^{(l)} \right\|_2^2 + 2( \boldsymbol{t}_n^{(l)} - \boldsymbol{t}_{n'}^{(l)})^T \notag\\
	&\quad\quad~~\times ( \boldsymbol{t}_n - \boldsymbol{t}_{n'})\ge D^2,~1\leq n \neq n' \leq N, \\
	&~f_k(\tilde{\boldsymbol{t}}^{\left(l\right)}) + \nabla f_k(\tilde{\boldsymbol{t}}^{\left(l\right)})^T (\tilde{\boldsymbol{t}} - \tilde{\boldsymbol{t}}^{\left(l\right)})+\notag\\ 
	&~~~~~~~\frac{\delta_k}{2}(\tilde{\boldsymbol{t}} - \tilde{\boldsymbol{t}}^{\left(l\right)})^T(\tilde{\boldsymbol{t}} - \tilde{\boldsymbol{t}}^{\left(l\right)})+\gamma_k\sigma_k^2 \le 0, \forall k.
\end{align}
\end{subequations} 
It can be observed that the problem (\ref{MA2}) is convex and can be solved by the existing optimization tools. The overall BCD algorithm is summarized in \textit{Algorithm 2}. 
\section{Numerical Results}
 \begin{figure}[t]
	\centering
	\includegraphics[width=0.28\textwidth]{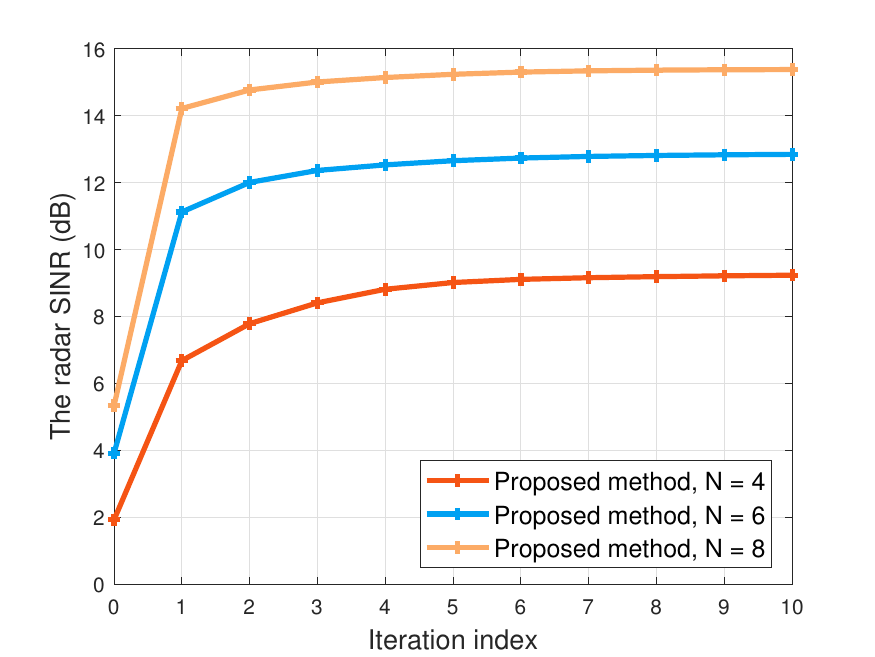}
	\captionsetup{font={normalsize},labelsep=period,singlelinecheck=off}
	\caption{Convergence behaviour of the BCD method.} 
	\label{convergence} 
\end{figure}%
In this section,  computer simulations are carried out to evaluate the performance of the proposed method.  We compare our scheme with four baseline schemes: \textbf{1)} \textbf{Fixed position antenna (FPA)}: The BS is equipped with  uniform planar arrays,  with $N$ transmitting/receiving antennas spaced between intervals of $\frac{\lambda}{2}$; \textbf{2)} \textbf{Random position antenna (RPA)}: The antennas at the BS are randomly distributed in the moving region under the constraint of the minimum distance $D$ between each other; \textbf{3)} \textbf{Random RIS}: The RIS phase shifts are generated randomly, following a uniform distribution within the range $[0,2\pi]$; \textbf{4)} \textbf{Greedy antenna selection (GAS)}: The moving regions are quantized into discrete locations spaced by $\frac{\lambda}{2}$. The greedy algorithm is employed for the optimization on antenna positions.

 In our simulation, we assume that the BS and the RIS are  located at (0, 0) m and (30, 5) m, respectively. The users are randomly distributed in  a circle centered at (30, 0) m with a radius of 3 m. The geometry channel model is employed for the communication links\cite{10318061}, where the numbers of transmitting and receiving paths are identical, i.e., $L^t_k = L^r_k = L = 4, \forall k$. Under this condition, the PRM for each  link  is diagonal, i.e., $\boldsymbol{\Sigma}_k = \mathrm{diag}\{\sigma_{k,1},\dots,\sigma_{k,L}\}$ with  $\sigma_{k,l} \sim \mathcal{CN}\left(0,\frac{c_k^2}{L}\right)$. Note that $c_k^2 = C_0d_k^{-\alpha}$ denotes the large-scale path loss, where $C_0 = -30~ \text{dB}$ is the expected average channel power gain at the reference distance of 1 m. The  path-loss exponents $\alpha$ for  the BS-RIS link, RIS-user link, and BS-target link are  given by 2.4, 2.8 and 2.6, respectively. Other parameters: $K = 3, N = 8, M = 32, Q = 2 ,\psi_0^e = 30^{\circ}, \psi_0^a = 45^{\circ}, \sigma_k^2 = \sigma_r^2 = \sigma^2= -80~\text{dBm},\gamma_k = \gamma = 10~\text{dB}, \lambda = 0.1~\mathrm{m}, D = \frac{\lambda}{2}, A = 2\lambda,  \mathcal{C}_t = \mathcal{C}_r=[-\frac{A}{2},\frac{A}{2}]\times[-\frac{A}{2},\frac{A}{2}].$ The AoDs and AoAs for the clutters are given by $\{\psi_1^e = 120^{\circ}, \psi_1^a = 90^{\circ}\}$ and $\{\psi_2^e = 135^{\circ}, \psi_2^a = 60^{\circ}\}$, respectively. 

 \begin{figure}[t]
 	\centering
 	\includegraphics[width=0.28\textwidth]{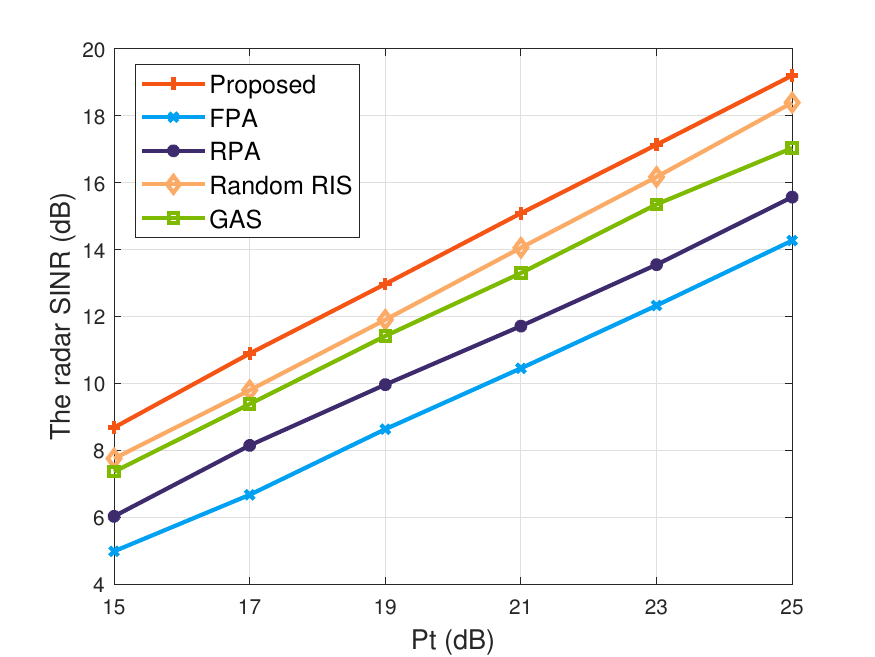}
 	\captionsetup{font={normalsize},labelsep=period,singlelinecheck=off}
 	\caption{The radar SINR versus transmission power $P_t$.} 
 	\label{Pt} 
 \end{figure}%
 
 \begin{figure}[t]
 	\centering
 	\includegraphics[width=0.28\textwidth]{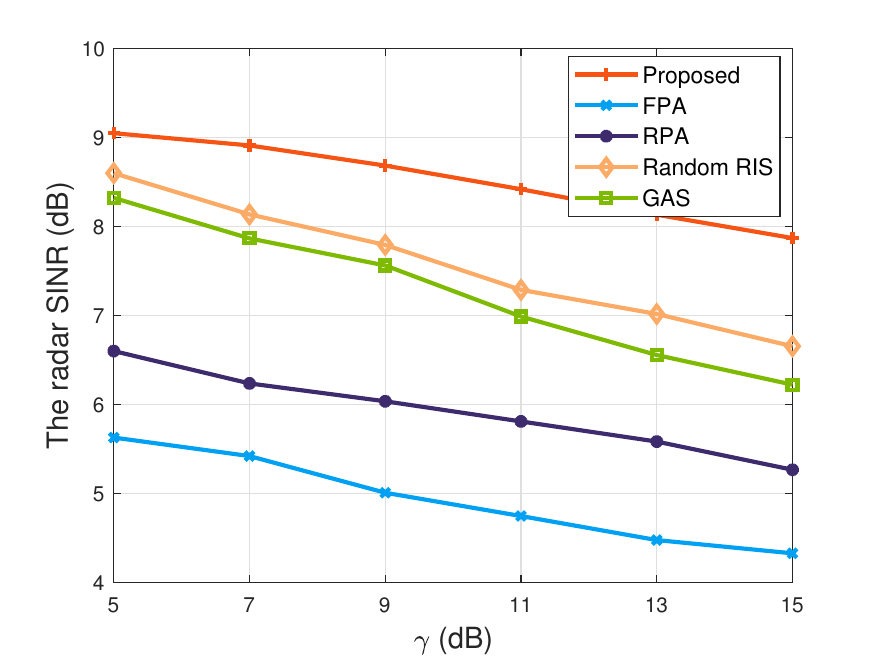}
 	\captionsetup{font={normalsize},labelsep=period,singlelinecheck=off}
 	\caption{The radar SINR versus communication QoS $\gamma$.} 
 	\label{QoS} 
 \end{figure}%
 In Fig. \ref{convergence}, we illustrate the radar SINR performance versus the iteration index. It can be observed that the  SINR   of configurations with varying numbers of antennas is monotonically increasing with the number of interation index. In most cases, less than 6 iterations are sufficient, verifying the effectiveness of the proposed method.
	
The impact of the maximum transmission power $P_t$ is plotted in Fig. \ref{Pt}. It can be observed that the proposed scheme can attain the highest radar SINR compared to the benchmark schemes, demonstrating the superiority of our algorithm. In particular, the proposed design can achieve about 5 dB radar SINR performance gain over the FPA scheme, which further validates the effectiveness of the MA technology.

In Fig. \ref{QoS}, we illustrate the relationship between the radar SINR performance and the communication QoS $\gamma$. It can be seen that the increasing $\gamma$ has minimal impact on the SINR of the proposed scheme. This is attributed to  the full spatial DoFs provided by MA, which allow simultaneous satisfaction of communication QoS and radar performance.

\section{Conclusion}
In this paper, we have proposed an optimization framework of the RIS-enhanced DFRC systems with movable antenna. A radar SINR maximization problem was formulated by jointly designing beamforming vectors, receiving filter, antenna positions and the RIS coefficients. Based on the FP principle, the BCD algorithm, along with SCA and penalty techniques, was invoked to solve this problem. Simulation results show that the proposed method can significantly improve the radar SINR, and achieve satisfactory trade-off between the radar and communication performance compared with existing benchmark schemes. Our method provides a generalized design framework for the RIS-MA-DFRC systems, and show the great potential of MA in RIS-aided DFRC.
\setlength{\footskip}{1in}
\balance
\bibliography{reference} 
\bibliographystyle{IEEEtran} 
%

\vfill
\end{document}